# Efficient hybrid search algorithm on ordered datasets

Adnan Saher Mohammed[a,*], Şahin Emrah Amrahov[b], Fatih V. Çelebi[c]

[a]*Ankara Yıldırım Beyazıt University, Graduate School of Natural Sciences, Computer Engineering Dept., Ankara, Turkey*
[b]*Ankara University, Faculty of Engineering, Computer Engineering Dept., Ankara, Turkey*
[c]*Ankara Yıldırım Beyazıt University, Faculty of Engineering and Natural Sciences, Computer Engineering Dept., Ankara, Turkey*

**Abstract**

The increase in the rate of data is much higher than the increase in the speed of computers, which results in a heavy emphasis on search algorithms in research literature. Searching an item in ordered list is an efficient operation in data processing. Binary and interpolation search algorithms commonly are used for searching ordered dataset in many applications.

In this paper, we present a hybrid algorithm to search ordered datasets based on the idea of interpolation and binary search. The proposed algorithm called Hybrid Search (HS), which is designed to work efficiently on unknown distributed ordered datasets, experimental results showed that our proposed algorithm has better performance when compared with other algorithms that use a similar approach.

## 1. Introduction

Searching algorithms are important and widely used in most computer systems. There are several search algorithms intended ordered datasets in literature. Binary search and interpolation search are the most used search algorithms which are utilized to search already sorted list. Both these algorithms are based on divide and conquer technique. Whereas, the dividing process depends on the way of choosing the cut index ($1 \leq cutindex \leq n$), where n is the size of the data set to be searched.

The main difference between binary search and interpolation search is the dividing method which is defined by way of computing the cut index. However, the dividing methods used by binary search and interpolation search have a valuable influence on the performance of these algorithms.

The complexity of the both algorithms were analyzed well in the literature. Asymptotically the average and the worst case complexity of binary search is $O(Log_2 n)$, which is independent of the distribution of keys [1]. Interpolation search was presented by Peterson in [2] with incorrect analysis [3], a very detailed description of the algorithm discussed along with program implementation by Gonnet Ph.D. thesis in [4] and the work [5]. The interpolation search average case and the lower band is ($log_2 log_2 n$) for uniform distributed key only [3, 5]. However, interpolation search shows slow performance for non-uniform distributed keys, up to $O(n)$ in the worst case. In other words, interpolation search is faster than binary search if some conditions are provided.

The author of [6] clarified two assumptions must be satisfied for an interpolation search to be practical. First, the access to a given key must be costly compared to a typical instruction. For example, in the case of external search when the array stored in external storage instead of internal memory. Second, the keys should be sorted and distributed uniformly. However, with the support of current powerful float point CPUs, interpolation search becomes faster than binary search even with internal memory search.

---

[*]Corresponding author
*Email addresses:* adnshr@gmail.com (Adnan Saher Mohammed), emrah@eng.ankara.edu.tr (Şahin Emrah Amrahov), fvcelebi@ybu.edu.tr (Fatih V. Çelebi)

On the other hand, author of [7] criticized interpolation search because it cannot stand over distribution changes. In literature, there exist some solutions to reduce the effect of this problem by proposing a hybrid approach that combines between binary search and interpolation search.

Our proposed algorithm is designed for a dataset with unknown distribution of keys. It is called Hybrid Search (HS) that experimentally shows a good performance for uniform and non-uniform distributed keys. In literature, there exist a similar solution which has been presented in [8], authors of the given algorithm called Adaptive Search (AS) proved their algorithm faster than an older algorithm was presented in [9]. However, in this work, we held a comparison study between our proposed algorithm and AS in [8] in section-5 .

This paper is organized as follows: section-2 presents the proposed algorithm and pseudo code, section-3 illustrates the complexity of the proposed algorithm , section-4 executes HS on a simple example array , section-5 compares our HS with AS, section- 6 gives experimental results and comparison , section-7 provides conclusions. Finally, you will find the important references.

## 2. The Proposed Hybrid Search (HS)

HS is designed to run competently on both uniform and non-uniform distributed sorted arrays. HS combines the basics of binary and interpolation search in an efficient manner to take the advantages of the both algorithms.

Assume HS looks to **X** among the sorted list **Array [Left .. Right]**, During each iteration, HS initially, calculates the estimated position by using the standard interpolation search method, then HS runs modified version of binary search on the remained part of the array which is expected holds the required key.

The next section provides the pseudo code of HS. By looking at the pseudo code more closely, we find HS firstly use interpolation technique to estimate the location of searched key. The calculated index is stored in the variable **Inter** (line No. 7). Then HS compares the required key **X** with the content of the determined location **Array[Inter]** in (lines No. 8 & 16) to decide whether the required key residents at the left or the right part respecting **Inter**. If **X** located on the left or the right portion, HS divides this segment into two halves by calculated the **Mid** variable in (lines No. 9 & 17). Hence, HS minimizes the search space by setting new values to the **Right** & **Left** variables.

After this process, the new search space for the next iteration has been minimized into half of the interpolated portion that was calculated by an interpolation method. This approach reduces the total number of iteration in all tested distributions. However, HS returns the location of the searched key in (lines No. 25 & 29). Considering that line 29 written to catch a case where the loop terminated and the element at **Array[Left]** not checked, for more detail follow the example in Figure 2. In the case of HS reaches the line No. 31, the algorithm ends with the unsuccessful search.

Figure 1 visually shows the division technique used by HS. Assume HS estimated the required key at index **Inter** (in Figure 1 ). If **X** greater than **Array[Inter]**, HS calculates the midpoint (**Mid**) of the range from **Inter** to **Right**. Then it selects one of these halves according to the value of **X**. Similarly, HS do the same procedures for the left portion ranged from **Left** to **Inter** when **X** less than **Array[Inter]**.

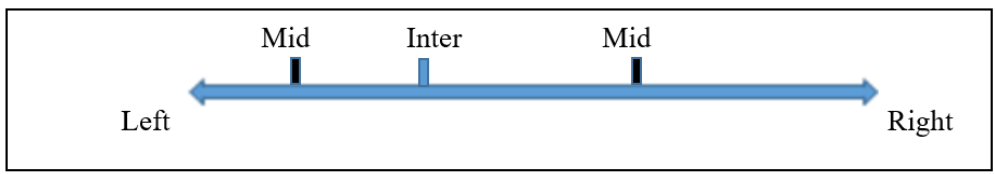

**Figure 1:** HS division technique explanation

```
Algorithm Hybrid Search (HS)
 1: procedure HS( array, left, right, X)
 2:     array is the array that required to search
 3:     left is the index of left most element in array
 4:     right is the index of right most element in array
 5:     X is the element that we search for
 6:     while left < right do
 7:         Inter ← left + (X−array[left])/(array[right]−array[left]) * (right − left)

 8:         if X > array[Inter] then                        ▷ to check the right portion of Inter
 9:             Mid ← (Inter+right)/2
10:             if X ≤ array[Mid] then
11:                 left ← Inter + 1
12:                 right ← Mid
13:             else
14:                 left ← Mid + 1
15:             end if
16:         else if X < array[Inter] then                   ▷ to check the left portion of Inter
17:             Mid ← (Inter+left)/2
18:             if X ≥ array[Mid] then
19:                 left ← Mid
20:                 right ← Inter − 1
21:             else
22:                 rigt ← Mid − 1
23:             end if
24:         else
25:             return Inter
26:         end if
27:     end while
28:     if X = array[left] then
29:         return left
30:     end if
31:     return −1                                           ▷ not found
32: end procedure
```

## 3. The Complexity of Hybrid Search

Although HS has higher iteration cost when compared with binary search and interpolation search due to its computations, HS showed less number of iteration compared with the binary search for all tested distributions and less than the interpolation search iterations for non-uniform distributions.

When HS runs on uniform distributed keys, the average case complexity of HS equal to the complexity of interpo-lation search, which is $O(log_2 log_2 n)$. That because HS starts with interpolation division technique in each iteration.

In the case of non-uniform distributed keys, the interpolation division technique works in conjugate with the binary division technique to reduce the number of iteration. However, HS worst case complexity still $O(log_2 n)$ in non-uniform distributions, because HS divides the interpolated segment into two halves in each iteration.

Experimental results explained that the number of iteration of HS is less than the half of the number of iterations of binary search in the average. Consequently, HS shows execution time much less than interpolation search and very close to the binary search for specified array sizes.

## 4. Hybrid Search Example

In this example, we randomly generated an ordered array that has location index starting from 0 to 34 with normally distributed keys. Then we searched all keys in this array using HS. Figure 2 explains the calculated variables (**Inter** & **Mid**) during each iteration in HS. Where bolded numbers on the figure, represent the returned location index by line No. 25 (return Inter) or line No. 29 (return Left) in the pseudo code section.

HS example given in Figure 2 shows iterations numbered between 1 and 3. The average iteration number equal to 2.31, while the average number of iterations for binary search on the same array is equal to 4.73, which is less than $log_2(35) = 5.12$.

By looking at Figure 2, we see HS found the first key and the last key (at index 0 and 34) in one iteration only, this is because the nature of interpolation search that used by HS. Base on this truth, assume we are looking for the item in the middle of the list (at index 17 in the example). HS certainly finds the middle item at the second iteration because this item becomes the last item after the first iteration. However, during each division process, Mid indexes to the one end of the new list and the other end indexed by **left** or **Inter** or **right** (see Figure 1), this method makes HS finds the required item exactly at the next iteration if it located at the end of the segment. Moreover, there is more chance to find the searched item in the next iteration if it located near to one end. Thus, this method enhances the performance of HS, especially for small or moderate array size.

| Index | Element/Key | Iteration 1 | | Iteration 2 | | Iteration 3 | |
|---|---|---|---|---|---|---|---|
| | | Inter | Mid | Inter | Mid | Inter | Mid |
| 0 | 6.983 | 0 | | | | | |
| 1 | 8.954 | 4 | 2 | 1 | | | |
| 2 | 11.74 | 11 | 6 | 4 | 2 | 2 | |
| 3 | 11.774 | 11 | 6 | 4 | 2 | 3 | |
| 4 | 12.192 | 12 | 6 | 4 | | | |
| 5 | 12.316 | 12 | 6 | 5 | | | |
| 6 | 12.682 | 13 | 7 | 6 | | | |
| 7 | 12.752 | 13 | 7 | 7 | | | |
| 8 | 12.939 | 14 | 7 | 7 | 10 | 8 | |
| 9 | 12.962 | 14 | 7 | 8 | 10 | 9 | |
| 10 | 13.42 | 15 | 8 | 9 | 11 | 10 | |
| 11 | 13.435 | 15 | 8 | 9 | 11 | 11 | |
| 12 | 13.99 | 16 | 8 | 12 | | | |
| 13 | 13.994 | 16 | 8 | 12 | 13 | left1 3 | |
| 14 | 14.446 | 17 | 9 | 15 | 12 | 14 | |
| 15 | 14.616 | 18 | 9 | 16 | 13 | 15 | |
| 16 | 14.634 | 18 | 9 | 16 | | | |
| 17 | 14.779 | 18 | 9 | 17 | | | |
| 18 | 14.961 | 18 | | | | | |
| 19 | 15.1 | 19 | | | | | |
| 20 | 15.101 | 19 | 26 | 20 | | | |
| 21 | 15.151 | 19 | 26 | 20 | 23 | 21 | |
| 22 | 15.602 | 20 | 27 | 22 | | | |
| 23 | 16.115 | 21 | 27 | 23 | | | |
| 24 | 16.131 | 21 | 27 | 23 | 25 | 24 | |
| 25 | 16.388 | 22 | 28 | 24 | 26 | 25 | |
| 26 | 17.053 | 23 | 28 | 27 | 26 | left2 6 | |
| 27 | 17.235 | 24 | 29 | 28 | 27 | left2 7 | |
| 28 | 17.35 | 24 | 29 | 28 | | | |
| 29 | 17.418 | 24 | 29 | 29 | | | |
| 30 | 17.575 | 25 | 29 | 30 | | | |
| 31 | 18.063 | 26 | 30 | 31 | | | |
| 32 | 18.925 | 28 | 31 | 32 | | | |
| 33 | 19.207 | 28 | 31 | 32 | 33 | left3 3 | |
| 34 | 21.374 | 34 | | | | | |

**Figure 2:** Hybrid Search Example

## 5. Hybrid Search and Adaptive Search Comparison

Fortunately, some authors have provided the source code of adaptive search in [10]. Adaptive search is the algorithm presented in [8]. Adaptive search is originally written in Java, simply we converted it to C++ code without altering the behavior of the algorithm to fit our test requirements. Table 1 explains C++ source code for adaptive search and HS. Although HS and adaptive search approximately revealed close iterations number in most cases, HS remarkably has less iteration cost in terms of comparison number and computation.

The adaptive search in Table 1 , initially, calculates values of **med** as the middle point and **next** by interpolation technique. Then, regarding the interpolated position (**next**), AS divides the selected portion (left or right segment) using binary search technique if and only if one of the following two conditions gets true (the first two *if* statements in AS):-

$$if ((key < a[next])\ \&\&\ ((next - bot) > med))$$
$$if ((key > a[next])\ \&\&\ ((top - next) > med))$$

Otherwise, AS continue as a normal interpolation search.

The main flaw in the dividing technique used in AS is that it may consume four comparison operations (inside the two mentioned *if* statements) without dividing the given segments into two halves. In regard, there is a probability of doing extra work in some iterations.

For instance, assume we are using AS to search for the element at the index 2 of the array in Table 1. In the first iteration bot=0 and top=34, AS calculates **med=17** and **next=11**. In this case, both *if* statements do not get true. However, the probability of occurrence of this case increases with a non-uniform distribution.

The advantages of our proposed algorithm over AS is that our algorithm uses both dividing techniques of binary search and interpolation search in each iteration using less number of comparisons. In Table 1, comments notations beside/after each *if* and *While* blocks indicate to the maximum number of comparisons in each iteration which consumed by CPU to execute these blocks. However, adaptive search shows 7 comparisons/iteration in the worst case and 5 comparisons/iteration in the best case. Whereas, HS shows 4 comparisons/iteration in the worst case and 3 comparisons/iteration in the best case. Consequently, in experimental performance test (section 6), HS shows better performance for different key distributions when compared with adaptive search in the same test environment.

Table 1: Adaptive Search And Hybrid Search Source Code Comparison

| Adaptive Search C++ code | Hybrid Search C++ code |
| --- | --- |
| ```cpp
int adaptiveSearch(double a[] ,
int bot , int top , double key)
{
  int  next, med;
  while ( bot < top)  // 1 Comps
  { med = (top + bot) / 2;
    next = (key - a[bot]) /(a[top]
         - a[bot]) * (top - bot) +
         bot;

    if ((key < a[next]) &&
        ((next - bot) > med))
        {// 3 Comps
          top = next - 1;
          next=(bot+top)/2;if (
    }
    (key > a[next]) && ((top -
       next) > med)) // 5 Comps
      {
        bot = next + 1;
        next = (bot + top) / 2;
      }
  if (key > a[next])   // 6 Comps
      bot = next + 1;
    else if (key < a[next])
         // 7 Comps
          top = next - 1;
         else
            return next;
  } // end while

  if (a[top] == key)
         return top;
  return -1;// not found
  }
``` | ```cpp
int HybridSearch (double a[], int
   left , int right, double X)
 {
  int Inter, Mid;
  while (left < right)   //1 Comps
   {
    Inter = left + (X - a[left]) * (
       right - left) /(a[right] - a[
       left]);
     if (x > a[Inter])  // 2 Comps
      {
        Mid = (Inter + right) / 2;
        if (x <= a[Mid])  // 3 Comps
         {
           left = Inter + 1;
           right = Mid;
         }
         else
            left = Mid + 1;
       }
     else if (x < a[Inter])  // 3 Comps
           {
             Mid = (Inter + left) /
                2;
             if (x >= a[Mid])
               {// 4 Comps
                left = Mid;
                right = Inter -1;
               }
             else
                right = Mid - 1;
            }
      else
         return Inter;
   } // end while
  if (x == a[left])
       return left;
  return -1; //not found
}
``` |

## 6. Results and Experimental Comparison

All algorithms in this study are implemented in C++ using NetBeans 8.0.2 IDE based on Cygwin compiler. The measurements are taken on a PC powered by 2.1 GHz Intel Core i7 processor with 6 GB PC-1066 DDR3 memory machine on Windows platform.

The experimental test has been done on empirical data that generated randomly using a C++ library [11]. For all



figures in this work, the execution time of the Binary search, interpolation search, adaptive search and hybrid search has been measured for 1000 repetitive runs of each algorithm on random array whereas the size of generated array ranged from 1000 to $23*10^6$ elements. 1000 search keys selected randomly among the elements of generated arrays, then these keys searched using all tested algorithms, this means, all algorithms tested for successful search only. The time readings have been plotted after applying smooth function in MATLAB software which is using moving average filter just to appear as smooth curve [12].

Measuring execution time is not very accurate operation in multitask/multithread operating system (OS) due to execution time variations caused by uncontrollable hardware and software factors. Even we run the same algorithm on the same data with the same environment [13, 14]. Furthermore, in this test, we run several algorithms on the multi-task operating system, and there is no guarantee from the operating system to distribute CPU time slices among tested algorithms fairly. We minimize this effect by configuring the operating system (Windows) on fixed CPU frequency and sets to give the highest process priority for our testing program.

All tested algorithms are fast and consume several milliseconds on used hardware. Therefore, it is hard to measure execution time precisely in one execution. To reduce the tolerant error caused by execution time variations and getting more accurate time measurement. All time measurements have done on repeated execution of the tested algorithms. Time measurement is taken for 1000 times of running an algorithm for each array size with randomly selected search key.

*6.1. Performance Test on Uniform Distribution*

In this test, we measured the execution time of all tested algorithms on the uniformly distributed sorted array. In general, the interpolation search is the fastest algorithm for this distribution, and the binary search is the slowest one. However, Figure 3 explains that HS performance very close to interpolation search in small sizes, and Figure 4 shows that HS delayed only up to 6% behind interpolation search for the array size $23*10^6$. Whereas, AS delayed about 50% for the same array size.

In brief, this test showed that HS faster than the binary search and AS for uniformly distributed array. And HS offered performance very close to the fastest search (interpolation search) in this distribution.

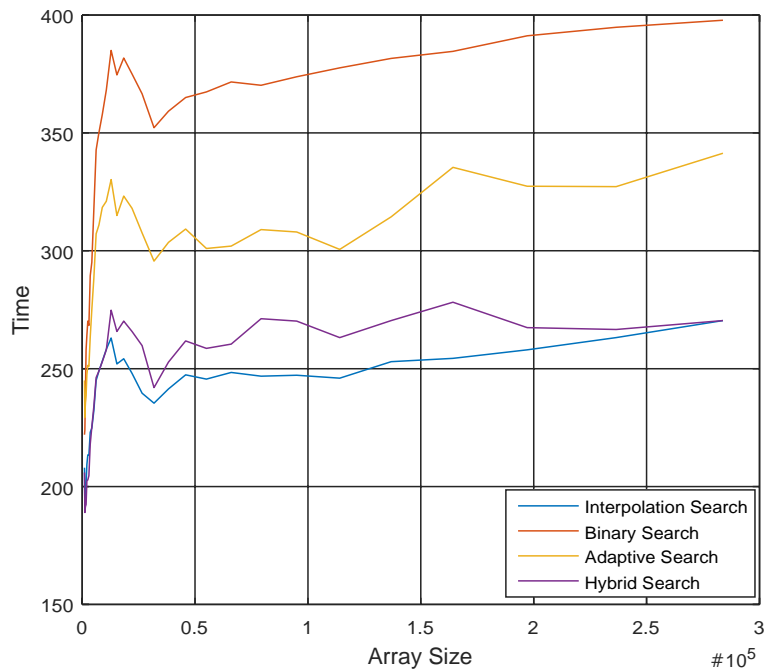

**Figure 3:** Execution-time of Interpolation, Binary, Adaptive and Hybrid Search on uniform distribution, small array size



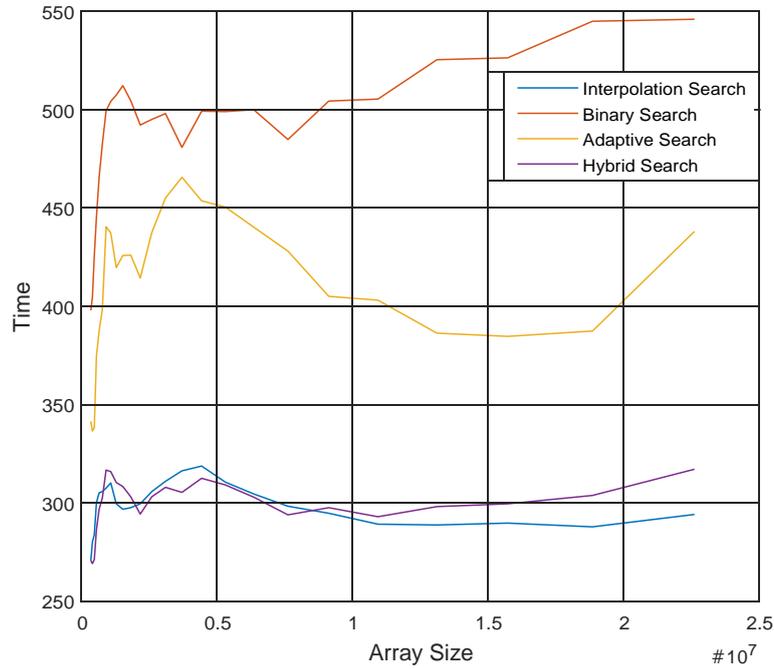

**Figure 4:** Execution-time of Interpolation, Binary, Adaptive and Hybrid Search on uniform distribution, large array size

## 6.2. Performance Test on Normal Distribution

In Figure 5, Obviously, interpolation search is the slowest algorithm due to its complexity in non-uniform distribution data set. HS shows better performance if compared with AS in a normal distribution. HS starts with an insignificant delay behind the binary search. Gradually, HS performance goes up to 12% slower than the binary search for array size less than 250,000 elements. However, AS goes about 50% slower than the binary search for the same size.

Figure 6 explains the performance of HS for large array size up to $23 \times 10^6$ elements. Even though, HS goes slower than the binary search up to 100%. Still, HS faster than AS in this distribution.

HS like interpolation search involving the keys in division (interpolation) process to expect the location of the searched key. Thus, HS performance depends on the distribution of the keys. On the other hand, HS use binary search division method which does not rely on the value of keys and their distribution. As a consequence, this technique makes the behaver of HS variant depending on the dataset length and the keys distribution.

When HS divides the given length into the half, gradually, the interpolation method becomes more effective because the distribution of data has been changed during previous division processes. That explains why HS showed better performance on small and moderate array size and less performance for large array size when compared with the binary search performance.



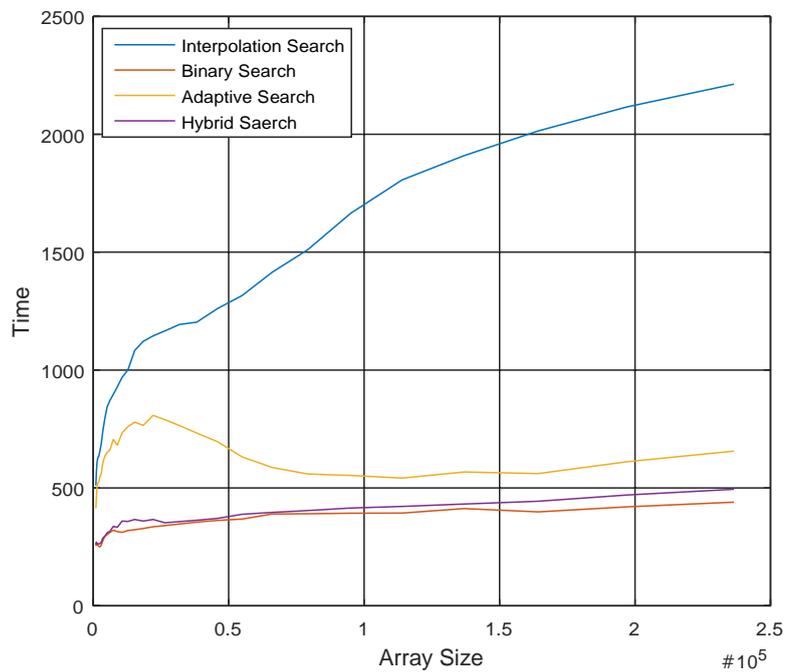

**Figure 5:** Execution-time of Interpolation, Binary, Adaptive and Hybrid Search on normal distribution, small array size

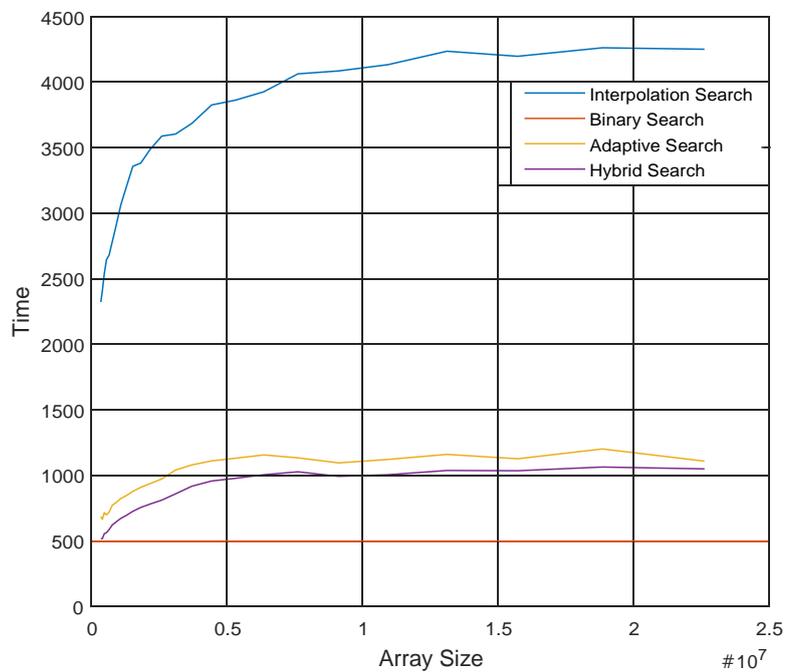

**Figure 6:** Execution-time of Interpolation, Binary, Adaptive and Hybrid Search on normal distribution, large array size



## 6.3. Performance Test on Exponential Distribution

Interpolation search shows even worse performance when compared with its performance on the normally distributed array in the previous section. Therefore, we excluded the interpolation search from this test to enhance the readability of the graph.

In Figure 7, HS displayed a very close performance comparing with the binary search for array size less than 300000 elements. Although, AS disclosed better performance when compared with its performance over the normally distributed array (see AS time values in Figure 5 & Figure 6). AS still slower than the binary search between 35%-75% for small and large array size respectively (see AS performance in Figure 7 & Figure 8).

Figure 8, explains that HS is the fastest algorithm for large array size in this distribution. In average, HS shows 10% faster than the binary search for array size up to $23 * 10^6$.

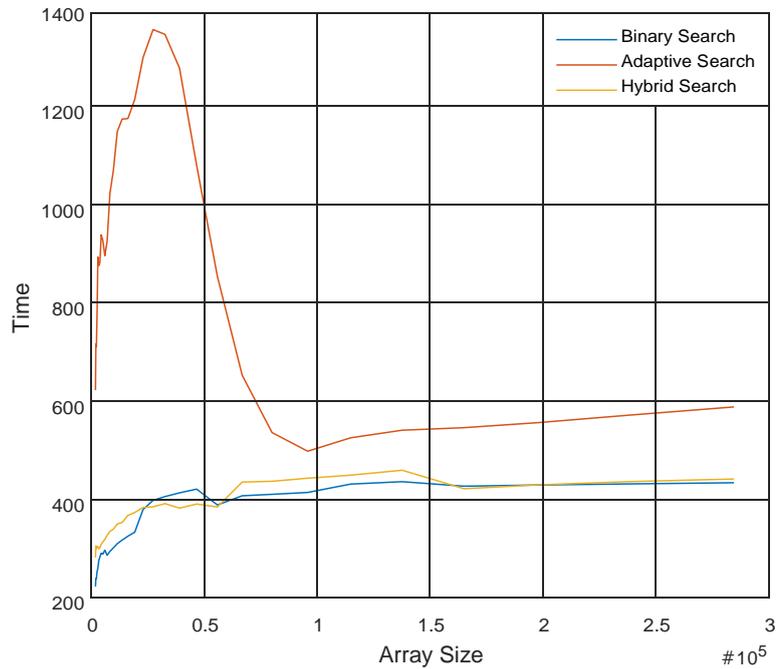

**Figure 7:** Execution-time of Interpolation, Binary, Adaptive and Hybrid Search on exponential distribution, small array size



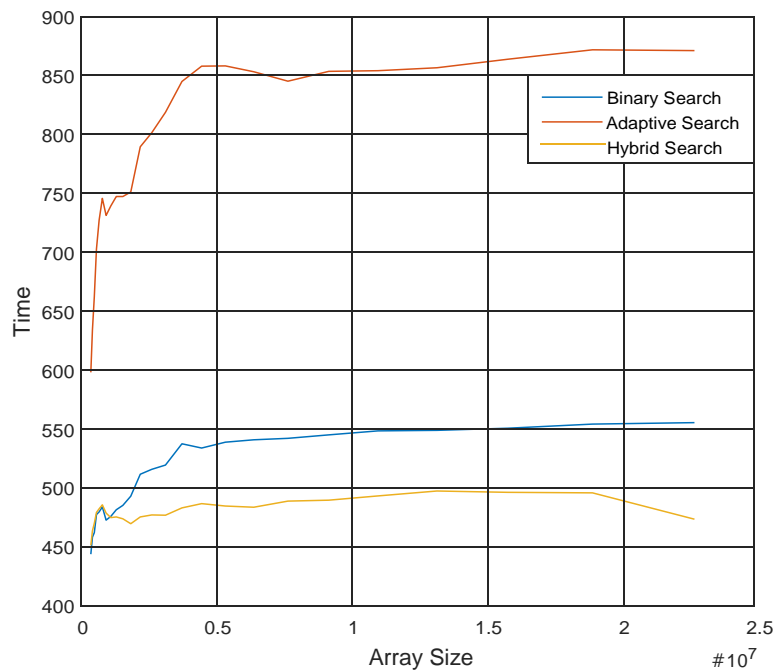

**Figure 8:** Execution-time of Interpolation, Binary, Adaptive and Hybrid Search on exponential distribution, large array size

## 7. Conclusions

In this paper, a hybrid searched algorithm has been presented, the proposed algorithm has been designed to use on unknown distributed sorted array which is based on the principle of binary and linear interpolation search. The proposed algorithm has been compared with three algorithms used for search on ordered datasets. The experimental results revealed that HS gives performance close to the fastest algorithm in each tested distribution, especially when HS runs on small or moderated array size. Hence, we recommend using HS as general search algorithm for unknown distributed ordered list within a certain situation that shows acceptable performance.

**Acknowledgments**

This research is partially supported by administration of governorate of Salahuddin and Tikrit university -Iraq (2245-6-2-2013).